\documentstyle[aps]{revtex}
\begin{document}
\draft
\title{ Constraints on inflation in the Einstein-Brans-Dicke frame}
\author{Yungui Gong \footnote{Email address: ygong@physics.utexas.edu}} 
\address{Physics Department, University of Texas at Austin, Austin, 
Texas 78712}
\maketitle
\begin{abstract}
The density perturbation during inflation seeds the large scale structure
of the Universe.
We consider both new inflation-type and chaotic inflation-type potentials
in the framework of 
Einstein-Brans-Dicke gravity. The density perturbation gives strong
constraints on the parameters in these potentials.
For both potentials, the constraints are not much different 
from those obtained in the original inflationary models
by using Einstein gravity.
\end{abstract}

\section{Introduction}

The successful explanation of cosmological puzzles, such as
the horizon, flatness, and monopole problems, is achieved by the 
inflationary scenarios. The inflationary scenarios also predict
the spectrum of the density perturbation which seeds the formation
of the large scale structure of the Universe. The basic assumption
is that, at the early times, the Universe experienced an accelerated expansion,
while the Hubble radius changed very slowly. Consequently the wavelength
of a quantum fluctuation soon exceeds the Hubble radius. The amplitude
of the fluctuation is frozen after the horizon crossing. After
the end of inflation, the Hubble radius increases faster than the 
scale factor; so the fluctuations eventually reenter the Hubble
radius during the radiation-dominated (RD) or matter-dominated (MD) eras.
The original model, the so-called ``old inflation'' model \cite{ag}, is based
on the first-order phase transition. It failed because of the ``graceful
exit'' problem. Soon after the ``new'' and ``chaotic'' inflationary
models were proposed
to solve this problem \cite{ncm}. These models use a simple scalar
field as the matter source. The scalar field (inflaton field) slow-rolls
down the potential during the inflationary phase. All the above models are
based on Einstein gravity. However, inflation may be driven by non-Einstein
gravity. ``Extended inflation'' 
employs Jordan-Brans-Dicke (JBD) gravity \cite{las}.
The introduction of the Brans-Dicke (BD) field slows down the inflation and
solves the graceful exit problem. But it was soon found that there was
a ``big bubble'' problem for the original extended inflation \cite{bubble}. 
The interaction between the BD field and the inflaton
field can change the spectrum of the density perturbation. The spectra
of density perturbations were analyzed in the extended new and chaotic 
inflationary models by several authors \cite{exdens}.
But the density perturbations given by those 
papers are not correct \cite{aasy}. The correct
density perturbation in BD inflation was given in \cite{aasy} \footnote{
Susperregi pointed out to me that the density perturbations obtained in
\cite{aasy} and \cite{msam} are consistent although they were
derived from two different approaches. I appreciate the comments made by
him.}. In this
paper, I use the slow-roll approximation and work with BD gravity 
in the Einstein frame (let us call it Einstein-Brans-Dicke gravity). There
are strong arguments identifying the Jordan frame \footnote{Both the Pauli
frame and the Einstein frame are used interchangeably in the literature.}
as the physical one.
The possibility of identifying the Einstein frame as the physical one
was first raised by Cho \cite{ymc}. Cho \cite{ymc} and Damour and
Nordtvedt \cite{d4} pointed out that only in the Einstein frame does the Pauli metric
represent the massless spin-2 graviton and the scalar field represent
the massless spin-0 field. In the Jordan frame the graviton is described
by both the Pauli metric tensor and the BD scalar field.
Cho also pointed out that in the compactification of Kaluza-Klein theory,
the physical metric must be identified as the Pauli metric because
of the wrong sign of the kinetic term of the scalar field in the Jordan
frame. In string theory, the dilaton field appears naturally. The Einstein
frame is greatly favored over the string frame although the string
frame is chose for the pre-big-bang cosmology.
For further discussions about 
the two frames, see \cite{gong} and references therein. Because the
dilaton field evolves very slowly during the RD and MD eras, we then assume that
the dilaton field at the end of inflation takes the same value as that
at present. By this assumption, we can fix the value of the inflaton
field at the beginning of inflation. We find that the results are different
from those in \cite{aasy}. Inflationary models based on general scalar-tensor
gravity in the Einstein frame were also
discussed in \cite{jgbdw}\footnote{The author thanks David Wands for pointing
out these references.} and \cite{exinf}.

The BD Lagrangian in Einstein frame is
\begin{equation}
\label{ebdlag}
{\cal L}= \sqrt{-g} \left[-\frac{1}{2\kappa^2}\Re
-\frac{1}{2}g^{\mu\nu}\partial_\mu \sigma \partial_\nu \sigma\right]
+{\cal L}_{m}(\psi, e^{-a\sigma}g_{\mu\nu}),
\end{equation}
where $\kappa^2=8\pi G,\  a=\beta\kappa$, and $\beta$ is a constant.
In this paper, I consider the simplest case, 
one simple scalar field as the matter source.
The generalization
to a more complicated multi-scalar-tensor gravity in Jordan
frame was discussed in \cite{d4}. For the cosmological models in the
context of general scalar-tensor
theory in Jordan frame, see \cite{jdb}. The matter Lagrangian is
given by
\begin{equation}
\label{matter}
{\cal L}_m(\psi, e^{-a\sigma}g_{\mu\nu})=-e^{-a\sigma}\sqrt{-g}
\left[{1\over 2}g^{ \mu \nu} \partial_ \mu \psi \partial_ \nu \psi 
+e^{-a\sigma}V(\psi)\right].
\end{equation}
The energy-momentum tensor for this matter source is
\begin{equation}
\label{ener}
T_{(m) \mu \nu}\equiv
{2\,\delta {\cal L}_m\over \sqrt{-g}\,\delta g^{\mu\nu}}=
e^{-a\sigma}\partial_\mu\psi\partial_\nu\psi+g_{\mu\nu}{\cal L}_m=
e^{-2\,a\sigma}\left[(\rho+p)\,U_\mu\,U_\nu + p\,g_{\mu\nu}\right],
\end{equation}
where 
$$U_\mu=-{\partial_\mu\psi\over \sqrt{-g^{\alpha\beta}\partial_\alpha\psi
\partial_\beta\psi}},\quad \dot{\psi}=\partial_\mu\psi U^\mu,$$
and $\rho={1\over 2}e^{a\sigma}\dot{\psi}^2+V(\psi)$,
$p={1\over 2}e^{a\sigma}\dot{\psi}^2-V(\psi)$.
With the assumption that the scalar field $\psi$ (inflaton field) is
spatially homogeneous, we can get the evolution equations of the
Universe from the actions (\ref{ebdlag}) and (\ref{matter}) based on the 
Robertson-Walker metric
\begin{mathletters}
\label{univ}
\begin{equation}
\label{scale}
H^2+{k \over R^2}={\kappa^2 \over 3}\left({\frac 1 2}\dot{\sigma}^2
+e^{-2a\sigma}\rho\right),
\end{equation}
\begin{equation}
\label{dila}
\ddot{\sigma}+3H\dot{\sigma}={\frac 1  2}a e^{-2a\sigma}(\rho-3p),
\end{equation}
\begin{equation}
\label{infla}
\ddot{\psi}+3H\dot{\psi}=-e^{-a\sigma}V'(\psi)+a\dot{\sigma}\dot{\psi},
\end{equation}
\end{mathletters}
where $V'(\psi)=dV(\psi)/d\psi$.

By the slow-roll approximations
\begin{equation}
\label{approx1}
{1\over 2}e^{2a\sigma}\dot{\sigma}^2\ll V(\psi),~~~
{1\over 2}e^{a\sigma}\dot{\psi}^2\ll V(\psi),~~~
\ddot{\sigma}\ll 3H\dot{\sigma},~~~
\ddot{\psi}\ll 3H\dot{\psi},
\end{equation}
Eqs. (\ref{scale})-(\ref{infla}) for a flat universe become
\begin{mathletters}
\label{approx}
\begin{equation}
\label{scale1}
H^2\approx {\kappa^2 \over 3}e^{-2a\sigma}V(\psi),
\end{equation}
\begin{equation}
\label{dilaton}
3H\dot{\sigma}\approx 2a e^{-2a\sigma}V(\psi),
\end{equation}
\begin{equation}
\label{inflaton}
3H\dot{\psi}\approx -e^{-a\sigma}V'(\psi).
\end{equation}
\end{mathletters}
The consistency conditions for the above approximations are
\begin{equation}
\label{consist}
\beta^2\ll {3\over 2},\quad {e^{a\sigma}V'\,^2(\psi)\over
6\kappa^2 V^2(\psi)}\ll 1,\quad
\biggl|{e^{a\sigma}V''(\psi)\over 3\kappa^2 V(\psi)}\biggr|\ll 1.
\end{equation}
From Eqs. (\ref{scale1})-(\ref{inflaton}), we get
\begin{mathletters}
\label{slou}
\begin{equation}
\label{gensolu}
\exp[a\sigma(\psi)]=2\kappa^2{\tilde m}_{Pl}^2\left({R(t)\over R(0)}\right)^{2\beta^2}
=2\kappa^2{\tilde m}_{Pl}^2\left[1-{\beta^2\over {\tilde m}_{Pl}^2}[g(\psi)-
g(\psi_i)]\right],
\end{equation}
\begin{equation}
\label{hubble}
H(\psi)=\sqrt{{\kappa^2\over 3}}e^{-a\sigma(\psi)}\sqrt{V(\psi)},
\end{equation}
\end{mathletters}
where $g(\psi)\equiv \int d\psi V(\psi)/V'(\psi)$ and
${\tilde m}_{Pl}<m_{Pl}$ is an arbitrary
integration constant corresponding to the effective Planck mass
at the beginning of inflation. Here I set the beginning
of inflation to be $t=0$. Remember that the variation of the 
dilaton field $\sigma$ is very small during
the MD era and the late times of the RD epoch.
Although the dilaton field may
have large changes during RD era for some initial values
\cite{gong},
we still assume that at the end of inflation the dilaton field becomes
$\sigma(t_e)\approx \ln(16\pi)/a$. Note that $e^{a\sigma}$ changes by a factor
of $(T_p/T_{EQ})^{\beta^2}\sim 10^{-5\beta^2}$ only from the matter-radiation
equality to the present.
Therefore, from Eq. (\ref{gensolu}), we get
\begin{equation}
\label{endcond}
N_{tot}=\ln {R(t_e)\over R(0)}={1\over 2\beta^2}\ln {m_{Pl}^2\over
{\tilde m}_{Pl}^2}.
\end{equation}
It is easy to get $N_{tot}\ge 65$. The conditions (\ref{consist}) tell
us that inflation ends when 
\begin{equation}
\label{endinf}
m^2_{Pl}V'\,^2(\psi(t_e))\approx 3\,V^2(\psi(t_e)),\qquad
{\rm or} ~~
|2\,m^2_{Pl}V''(\psi(t_e))|\approx 3\,V(\psi(t_e)).
\end{equation}

The density fluctuation due to primordially adiabatic fluctuation
is given by \cite{aasy}
\begin{eqnarray}
\label{dens}
{\delta\rho\over \rho}&=&f{H^2\over 2\pi}\left[{e^{3a\sigma/2}\over 16\pi\,
\dot{\psi}}
+{1-e^{a\sigma}/(16\pi)\over \dot{\sigma}}\right]_{t_k}  \nonumber \\
&=&
{f{H\over 2\pi}\left[-{e^{a\sigma/2}\kappa^2 V(\psi)\over 16\pi\,V'(\psi)}
+{\kappa\over 2\beta}[1-e^{a\sigma}/(16\pi)]\right]_{t_k}},
\end{eqnarray}
where $f$ equals $4/9$ if the fluctuations reenter the horizon
during the RD era or $2/5$ if the fluctuations reenter the horizon during
the MD era, and $t_k$ is the time at the horizon crossing during
the inflationary epoch. For convenience, we use the number of $e$-foldings,
$N_k$, before the end of inflation, instead of $t_k$. As usual, we take
$N_k\approx 55-60$. In terms of $N_k$, we have
$$\exp[a\sigma(t_k)]\approx 16\pi\exp(-2\beta^2 N_k).$$
In order to go further, we need to specify the form of the potential.

\section{New Inflation-Type Potential}

In this section, we take the Coleman-Weinberg potential \cite{cw},
\begin{equation}
\label{potent}
V(\psi)={B\,\eta^4\over 2}+B\,\psi^4\,\left[\ln(\psi^2/\eta^2)
-{1\over 2}\right],
\end{equation}
where $B\approx 10^{-3}$, $\eta\approx 2\times 10^{15}$ GeV.
The above potential can be well approximated by 
\begin{equation}
\label{potent1}
V(\psi)\approx {B\,\eta^4\over 2}-{\lambda\over 4}\psi^4
=V_0-{\lambda\over 4}\psi^4,
\end{equation}
where $V_0=B\,\eta^4/2$ and $\lambda=|4\,B\,\ln(\psi^2/\eta^2)|$
is approximately a constant. 
From the potential (\ref{potent1}), we know that
inflation will end when
$$ m^2_{Pl}V'\,^2(\psi(t_e))\approx 3V^2(\psi(t_e)), \qquad
{\rm if} ~~{1\over 4}\lambda\psi^4(t_e)>{3\over 5}V_0,$$
$$|2\,m^2_{Pl}V''(\psi(t_e))|\approx 3V(\psi(t_e)), \qquad 
{\rm if} ~~{1\over 4}\lambda\psi^4(t_e)\le {3\over 5}V_0.$$
The inflaton field at the end of inflation takes the value
\begin{equation}
\label{inflatone}
\psi_f^2=\psi^2(t_e)\approx
-4m^2_{Pl}+{2\over \lambda}\sqrt{\lambda V_0+4\lambda^2 m^4_{Pl}}
\approx {V_0\over 2\lambda m^2_{Pl}}.
\end{equation}
In the last step, we use the result $V_0\ll 4\lambda m^4_{Pl}$. Because
$\lambda \psi^4_f/4\ll V_0$, it is well justified to approximate
$V(\psi)\approx V_0$ during inflation. By using this approximation,
we find the time evolutions of the scale factor, dilaton field, and
inflaton field as
$$\exp[a\sigma(t)]=2\kappa^2\,{\tilde m}^2_{Pl}(1+2\,\beta^2\,H_v\,t),$$
$$R(t)=R(0)(1+2\,\beta^2\,H_v\,t)^{1/2\,\beta^2},$$
$${1\over \psi^2}={1\over \psi^2_i}-
{\lambda\over 3\kappa^2{\tilde m}^2_{Pl}H_v}t=
{1\over \psi^2_i}-{2\,\lambda\over\sqrt{3\,V_0}\kappa}t,
$$
where $H_v=\sqrt{V_0}/2\sqrt{3}\kappa\,{\tilde m}^2_{Pl}$ is the Hubble
parameter at the beginning of inflation.
The function $g(\psi)$ is
\begin{equation}
\label{auxn}
g(\psi)=\int d\psi {V(\psi)\over V'(\psi)}\approx
 -\int d\psi {V_0\over \lambda\,\psi^3}={V_0\over 2\lambda}\psi^{-2}.
\end{equation}
Substituting Eqs. (\ref{inflatone}) and (\ref{auxn}) back 
into Eq. (\ref{gensolu}), we find that
\begin{mathletters}
\label{newsolu}
\begin{equation}
\label{inflatoni}
\psi_i^2={\beta^2\,V_0\over 2\lambda [(1+\beta^2)m^2_{Pl}
-{\tilde m}^2_{Pl}]}\approx \beta^2\,\psi_f,
\end{equation}
\begin{equation}
\label{infln}
\psi^2(t_k)={\beta^2 V_0\over 2\lambda m^2_{Pl}[1+\beta^2
-\exp(-2\beta^2\,N_k)]}.
\end{equation}
\end{mathletters}
Therefore the density perturbation is
\begin{mathletters}
\label{newdensp}
\begin{equation}
H(t_k)={1\over 2\sqrt{24\pi}}
e^{2\beta^2 N_k}\sqrt{{V_0\over m^2_{Pl}}}, 
\end{equation}
\begin{equation}
\label{densn}
{\delta \rho\over \rho}={1\over 5\sqrt{3}\pi\,\beta^3}e^{\beta^2 N_k}\left[
1+\beta^2-e^{-2\beta^2\,N_k}\right]^{3/2}\sqrt{\lambda}
+{\sqrt{3}\over 60\pi\beta}(e^{2\beta^2\,N_k}-1){\sqrt{V_0}\over
m^2_{Pl}}.
\end{equation}
\end{mathletters}
Note that our formula is different from that in \cite{aasy}. 
Take $N_k=60$, $\beta^2=0.001$ \cite{yzg}, then we get 
$${\delta \rho\over \rho}=47.6\sqrt{\lambda}+0.04{\sqrt{V_0}\over
m^2_{Pl}}.$$
The second term in the right hand side is much less than $10^{-5}$.
Therefore, if we require
$\delta \rho/\rho<10^{-5}$, we must have
\begin{equation}
\label{paran}
\lambda<4.4\times 10^{-14}.
\end{equation}
Note that the minimum value of the coefficient 
before $\sqrt{\lambda}$ in Eq. (\ref{densn})
is about 38.4 when $\beta\approx 0.13$.
If $\beta^2\ge 0.04$, the coefficient increases quickly and
it requires smaller $\lambda$ to get the right density perturbation.
In \cite{bmy},
the author did not get the constraint on $\lambda$ because they
considered the fluctuation from the dilaton field only. However, it is 
clear form Eq. (\ref{densn}) that the fluctuation from the
inflaton field is larger than that from the dilaton field. In fact, our result
is easily understood. Note that the fluctuation due to the inflaton field
is
\begin{equation}
\label{flucinf}
\delta\psi(t)={e^{a\sigma(t)/2}H(t)\over 2\pi}\le {1\over 2\pi}
{\sqrt{16\pi}\over 2\sqrt{24\pi}}\sqrt{{V_0\over {\tilde m}^2_{Pl}}}
={1\over 2\pi}\sqrt{{V_0\over 6{\tilde m}^2_{Pl}}}.
\end{equation}
Here we use $\psi(t)>\psi_i$ and $\psi_i$ is given in Eq. (\ref{inflatoni}).
To make our semiclassical discussions valid, we must require that
the classical value of the inflaton field be larger than its
quantum fluctuation.  So we have
\begin{equation}
\label{paran1}
\lambda<{12\pi^2\,\beta^2 {\tilde m}^2_{Pl}\over m^2_{Pl}}.
\end{equation}
The smallness of the value of $\lambda$ cannot be saved.
In \cite{am}, Matacz derived a different formula for the scalar quantum
fluctuation $\delta\psi$ based on stochastic approach and gave a constraint
$\lambda\sim 10^{-5}$. Perhaps that is one way to avoid the fine-tuning
problem.

\section{Chaotic Type Potential}

For chaotic inflation, the potential takes the power law type
\begin{equation}
\label{potentc}
V(\psi)={\lambda_n\over n}\psi^n,
\end{equation}
where $n$ is an even integer. In this case, the function $g(\psi)$ is
\begin{equation}
\label{auxc}
g(\psi)=\int d\psi {V(\psi)\over V'(\psi)}=
\int d\psi {\psi\over n}={\psi^2\over 2n}.
\end{equation}
Because $n\ge 2$, the end of inflation will happen when
\begin{mathletters}
\label{endinfl}
\begin{equation}
|2\,m^2_{Pl}V''(\psi(t_e))|\approx 3V(\psi(t_e)),
\end{equation}
\begin{equation}
\label{cinfe}
\psi_f^2={2n(n-1)\over 3}m^2_{Pl}.
\end{equation}
\end{mathletters}
Combining Eqs. (\ref{gensolu}), (\ref{auxc}), and (\ref{cinfe}), we find
\begin{mathletters}
\label{tpsi}
\begin{equation}
\label{cinfi}
\psi_i^2=\left({2n\over \beta^2}+{2n(n-1)\over 3}\right)m^2_{Pl}
-{2n\over \beta^2}{\tilde m}^2_{Pl}\approx 
\left({2n\over \beta^2}+{2n(n-1)\over 3}\right)m^2_{Pl},
\end{equation}
\begin{equation}
\label{inflc}
\psi^2(t_k)={2n\over \beta^2}\left(1+{(n-1)\beta^2\over 3}-\exp(-2\beta^2\,N_k)
\right)m^2_{Pl}.
\end{equation}
\end{mathletters}
In terms of $N_k$, we find
\begin{mathletters}
\label{chaodens}
\begin{equation}
H(t_k)={1\over 2\sqrt{24\pi}}e^{2\beta^2 N_k}\left[
{2n\over \beta^2}\left(1+{(n-1)\beta^2\over 3}-\exp(-2\beta^2\,N_k)
\right)m^2_{Pl}\right]^{n/4}\sqrt{{\lambda_n}\over n m^2_{Pl}},
\end{equation}
\begin{equation}
\label{densc}
{\delta \rho\over \rho}={1\over 5\pi}\sqrt{{\lambda_n\over n}}m_{Pl}^{n/2-2}
[h(N_k)]^{n/4}\left|
-{1\over 2n\sqrt{6}}e^{\beta^2 N_k}\sqrt{h(N_k)}
+{1\over 4\sqrt{3}\beta}(e^{2\beta^2 N_k}-1)\right|,
\end{equation}
\end{mathletters}
where 
$$h(N_k)={2n\over \beta^2}\left(1+{(n-1)\beta^2\over 3}-\exp(-2\beta^2\,N_k)
\right).$$

For $n=2$, we have
$${\delta \rho\over \rho}=1.7{\sqrt{\lambda_2}\over m_{Pl}}.$$
So the bounds on the anisotropy of the microwave background give
\begin{equation}
\label{parac2}
\lambda_2<4\times 10^{-11} m^2_{Pl}.
\end{equation}

For $n=4$, we have
$${\delta \rho\over \rho}=30.7\sqrt{\lambda_4}.$$
In order to get the small density perturbation, we require
\begin{equation}
\label{parac4}
\lambda_4<1\times 10^{-13}.
\end{equation}

At last, let us look at the exponential potential
$V(\psi)=V_0\exp(-\psi/\psi_0)$ \cite{dsjm}. In this case, both $V'(\psi)/V(\psi)$
and $V''(\psi)/V(\psi)$ are constants. Then the slow-roll condition
cannot determine when the inflation will end under the assumption that
$\exp(a\sigma)=16\pi$ at the end of inflation. Therefore, this kind of
potential is not workable in our concern. In this model, we must consider
the interaction between the inflaton field and other fields to let
the Universe exit from the inflationary epoch. In \cite{msam}, the authors
considered the exponential potential in extended inflation. They
used the consistency conditions (\ref{consist}) to give the value
of the BD field at the end of inflation. Since we know the evolution
of the BD field during the RD and MD eras, it may be a problem to match
the value of the BD field from the end of inflation to the present.
We would like to say a few more words about the difference between our
results and those in \cite{aasy}. As a result of the assumption that the dilaton
field takes the same value at the end of inflation and the present,
we can fix the value of the
inflaton field at both the beginning and the end of inflation.
If we take the approximation $\exp(-2\beta^2 N_k)\approx 1-2\beta^2 N_k$,
then our results (\ref{densn}) and (\ref{densc}) are similar to those
in \cite{aasy}. For the choices of $\beta^2$ and $N_k$ in
this paper, we can take
this approximation. That is why our numerical values do not differ much
from those in \cite{aasy}.
Instead of thinking the physical frame to be the Jordan frame, 
we work in the Einstein frame. 

\acknowledgments

The author would like to thank Professor Yuval 
Ne'eman for his helpful comments.

\end{document}